\documentclass{optica-article}

\journal{opticajournal} 

\articletype{Research Article}

\usepackage{lineno}
\usepackage{float}
\usepackage{xcolor}

\begin{document}

\title{Conditional Recurrent Neural Networks for broad applications in nonlinear optics}

\author{Simone Lauria\authormark{*}, Mohammed F. Saleh}

\address{SUPA, Institute of Photonics and Quantum Sciences, Heriot-Watt University, EH14 4AS Edinburgh, UK}

\email{\authormark{*}sl196@hw.ac.uk} 


\begin{abstract*} 
We present a novel implementation of conditional Long Short-Term Memory Recurrent Neural Networks that successfully predict the spectral evolution  of a pulse in nonlinear periodically-poled waveguides. The developed networks offer large flexibility by allowing the propagation of optical pulses with ranges of energies and temporal widths in waveguides with different poling periods. The results show very high agreement with the traditional numerical models. Moreover, we are able to use a single network to calculate both the real and imaginary parts of the  pulse complex envelope, allowing for successfully retrieving the pulse temporal and spectral evolution using the same network. 

\end{abstract*}

\section{Introduction - Machine Learning}
Machine learning is a branch of artificial intelligence and computer science that involves the implementation of numerical algorithms, which autonomously learn to successfully make predictions in various fields \cite{GeneralML, ML_trends}.  It has been extensively applied in image recognition and classification \cite{classification1, calssification2}, natural language processing \cite{NLP}, time-series prediction \cite{timeseries}, cybersecurity \cite{cybersec}, healthcare \cite{healthcare}, autonomous vehicle control \cite{vehicle} and neuroscience research \cite{neuroscience}. 
In the recent years, machine learning techniques have also been developed for different applications in optics \cite{NLML1,NLML2, NLML3}, such as inverse design of photonic structures \cite{inverseDesign}, and optical microscopy \cite{microscopy}.

Nonlinear optical pulse propagation presents another optimal field for harnessing the capabilities of neural networks and machine learning. A number of different architectures and implementations have been proposed in recent years, ranging from physics-informed solutions  that rely on knowledge by the network of the governing equations \cite{physicsNN}, to fully model-free, data-driven methods that can, for instance, predict the outcomes of the numerical solution of the Nonlinear Schr\"{o}dinger equation \cite{gentyFFNN,gentyRNN, ConvNN}, or improve  numerical simulations of optical Rogue-waves \cite{NNaidNLSE}. Optical pulse propagation in a nonlinear medium depends on the parameters of the input pump as well as the geometry of the medium. The latter can drastically affect the pulse dynamics through modifying the dispersion profile, and the nonlinearity strength. Distinguishing between the nonlinear pulse propagation in a range of different structures would often require training multiple neural networks, each for a certain structure.

In this work, we develop a fully data-driven approach using conditional recurrent neural networks (RNN) \cite{rnn_design} to predict the outcomes of nonlinear pulse propagation for a range of input-pulse parameters in different waveguide structures that simultaneously exhibit second- and third-order nonlinearities. The developed approach also allows for the training of a unified network to compute both the spectral and temporal evolution of a pulse via assigning the real and imaginary parts of the pulse complex-envelope as the network condition. We present in this paper the architecture of the conditional RNN followed by different examples that demonstrate its potential in successfully simulating the dynamics of ultrashort pulses in nano-photonic periodically-poled lithium-niobate waveguides.

The paper is organised as follows: In Sec. 2, we present the numerical model, traditional RNNs and the conditioning of sequential data.  Two trained networks based-on a single conditional recurrent architecture are discussed in Sec. 3. A trained network based on a dual conditional architecture, with two separate conditioning parameters, is explained in Sec. 4. Finally, our conclusion together with a comparison between  the performances of the different conditional RNNs and the numerical model are presented in Sec. 5. 

\section{Data driven RNNs}
\subsection{Numerical Model}
The complex dynamics of the pulse electric field in periodically poled waveguides with simultaneous second- and third-order nonlinearities can be  simulated with large accuracy using the unidirectional pulse propagation equation (UPPE) \cite{UPPE,lauria},

\begin{equation}
\frac{\partial \Tilde{E}}{\partial z} = i [\beta (\omega) - \beta_1 \omega ] \Tilde{E} + i\frac{\omega ^2}{2 c^2  \beta (\omega)} \mathcal{F} \{  \chi^{(2)}(z) E^2(z, t) +  \chi^{(3)} E^3(z, t) \},\label{eq1}
\end{equation}
where $z$ is the propagation axis, $\omega$ is the angular frequency, $\tilde{E}(z,\omega) = \mathcal{F} \{E (z, t) \}$ is the spectral electric field, $t$ is the time in a reference frame moving with the pulse group velocity, $\mathcal{F}$ is the Fourier transform, $\beta (\omega)$ is the full dispersion, $\beta _1$ is the first-order dispersion coefficient, $c$ is the speed of light in vacuum, and $\chi^{(2)}$ and $\chi^{(3)}$ are the second- and third-order nonlinear coefficients, respectively. The pulse complex envelope can be extracted using $\mathcal{A}( (z, t)=\mathcal{E} (z, t)e^{i [\omega_0 t-\beta (\omega_0)z]}$, with $\omega_0$  the pulse central frequency, $\mathcal{E}$ the analytical signal given by $\mathcal{E}(z, t)=E(z, t)-i\mathcal{H}\left[E(z, t)\right]$, and  $\mathcal{H}$ the Hilbert transform \cite{Conforti13}.

The UPPE model is solved via implementing the split-step Fourier method \cite{Agrawal07}, over a longitudinal stepsize in the range of $\Delta z \leq 0.2 \,\mu$m. For certain poling periods, we found that a very small stepsize is required to prevent the model from divergence. Hence, to optimise the performance of the model, an adaptive algorithm that decreases $\Delta z$ only when required is applied. Although the solution is largely accurate, it is very time consuming and computationally demanding due to the extremely fine stepsize needed, together with the fine spectral resolution necessary for modelling the electric field.

\subsection{RNNs with LSTM layers}
\begin{figure}
    \centering
    \includegraphics[width=0.8\textwidth]{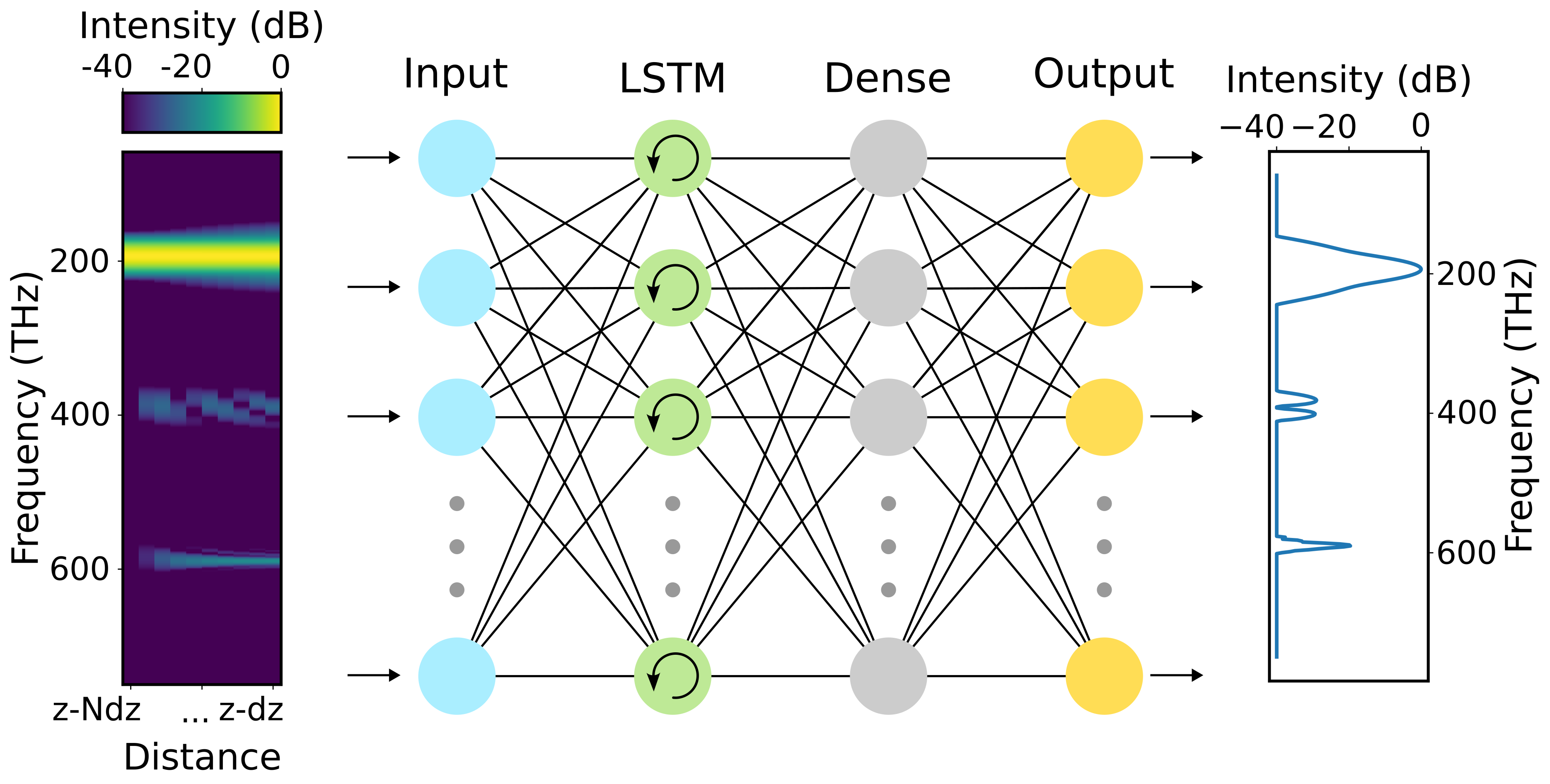}
    \caption{Standard implementation of a LSTM network for predicting the spectral evolution of a pulse. 
    $N$ steps of the propagation,  shown in a color plot (left) are fed to the network (middle) to predict the next step (right).}
    \label{fig:LSTM SIMPLE}
\end{figure}

Using recurrent neural networks can drastically reduce the simulation time of numerical solutions. RNNs are ideal for learning sequential data, due to the feedback loops, or connections within their layers \cite{rnns_time,rnn_design}, that provide an internal memory of past observations. In particular, RNNs with long short-term memory (LSTM) \cite{lstm} layers can deal with longer sequences, where the sequential dependence of the data spans over a large number of observations, since these types of network avoid vanishing gradients problems \cite{vanish}. The standard implementation of a LSTM recurrent network consists of feeding $N$ sequential inputs at equally-spaced $\Delta z$ positions to the network, to predict the output at the next step. For instance,  the pulse spectral profiles at $N$ positions are fed to a neural network to calculate the output at the position $N+1$, as shown in  the simplified diagram of a LSTM network, in Fig. \ref{fig:LSTM SIMPLE}.Practically,  the network consists of more LSTM and Dense layers with varying number of nodes. A limitation of this architecture is being structure-specific, since the sequential data learned by the network lacks any information about the structure, which can dramatically  affect the output.

\subsection{Conditional RNN}
\begin{figure}
    \centering
    \includegraphics[width=0.8\textwidth]{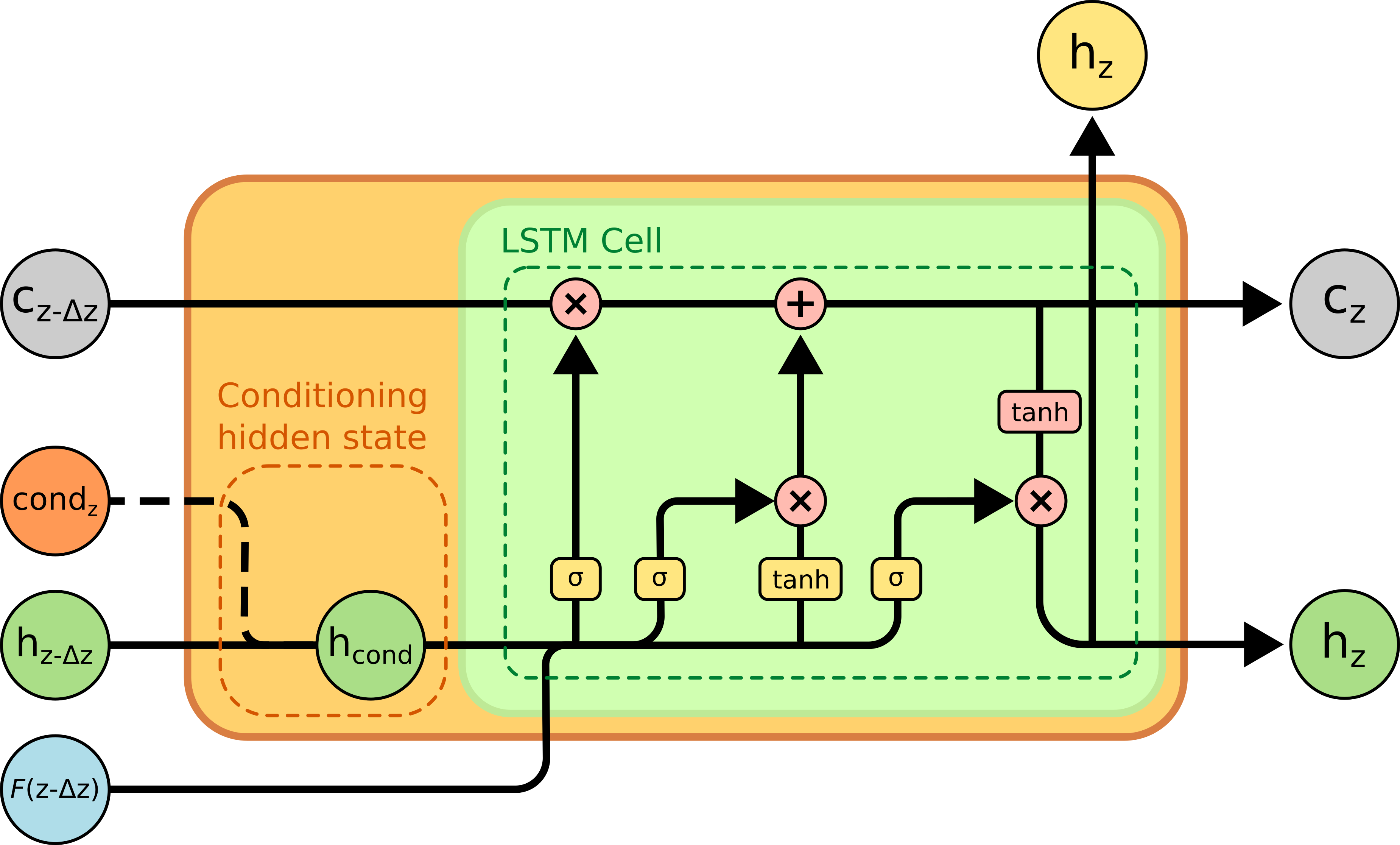}
    \caption{Diagram of ConditionalRecurrent (in orange) around a LSTM cell (green).
    The wrapper embeds the condition cond$_z$ within the hidden state of the LSTM cell, h$_{z-\Delta z}$, feeding the conditioned hidden state h$_{\mathrm{cond}}$ to the cell. The cell state c$_{z-\Delta z}$ and the input $F(z-\Delta z)$ are not affected.}
    \label{fig:cond}
\end{figure}

Feeding additional data is therefore crucial for predicting different dynamics that are affected by non-sequential data, such as the waveguide parameters. Such data can be fed to a RNN by appending it to the sequence, as an additional value or vector within the input matrix. This approach relies on the network learning autonomously that the additional information is not part of the sequential data, but rather a conditioning parameter, which may lead to very long and possibly unsuccessful training. Alternatively, the conditioning of the prediction can be performed \textit{a posteriori}, by concatenating the RNN with another form of neural network. However, this method is ineffective in capturing a wide range of dynamics. An ideal approach is instead to condition the hidden state \cite{lstm} of the recurrent or LSTM cell, as it is fed to the cell itself. This solution has been integrated in the form of a Keras wrapper for recurrent layers, the ConditionalRecurrent \cite{remy}, shown in Fig. \ref{fig:cond}.  The wrapper acts on the input hidden state of every cell, allowing for the architecture to be implemented within a model similarly to a standard LSTM cell. 

\section{Single conditional RNN}
\begin{figure}
    \centering
    \includegraphics[width=0.9\textwidth]{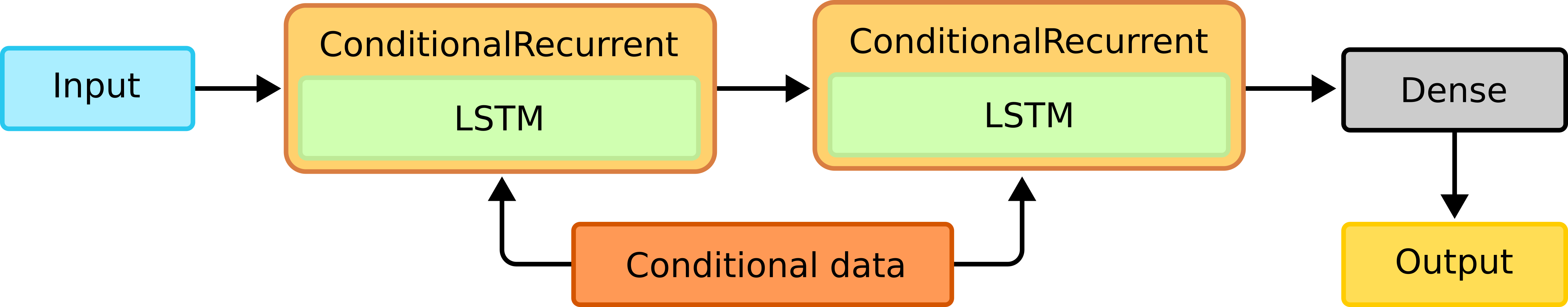}
    \caption{Architecture of a single-condition RNN, where the conditional data is fed twice to the network, while the sequential data is only fed once.}
    \label{fig:Single}
\end{figure}

The architecture of the conditional RNN with a single condition is shown in Fig. \ref{fig:Single}. The model has been designed with the Keras functional API \cite{keras} that allows for a non-sequential model, in account of the multiple inputs, i.e. the sequential and conditional data. To correctly initialise the hidden states, the conditional data is fed twice to the network, initially with the sequential data, then with the conditioned sequential data output from the first LSTM layer. This has been proven to be a good compromise between the quality of the prediction and the associated speed.

\subsection{Uniformly and linearly chirped poled waveguides}
We have trained the RNN to predict the spectral pulse evolution in different uniformly-poled thin-film lithium niobate waveguides with a trapezoidal core, 950 nm top width, 340 nm etching height, 800 nm thickness, 60 degree side angle, 4.56 mm length, and silica as a bottom cladding. The pulse central wavelength is fixed at 1550 nm, whereas the pulse full-width-half-maximum (FWHM) and energy are varied over the ranges 25 -- 30 fs and 3 -- 5 pJ, respectively. This is equivalent to dispersion and nonlinearity lengths in the ranges 5.27--7.59 mm and 0.82--1.63 mm respectively. The uniform poling of the waveguide is varied over the range 3.2 -- 4.4 $\mu$m. The second and third-order coefficients are assumed as $\chi ^{(2)}=26$ pm/V, $\chi ^{(3)}=3417$ pm$^2$/V$^2$, respectively.

The network is trained with the simulations generated by the UPPE model with the poling period as the conditional parameter. Ten sequential steps of the spectral profile are fed together with the poling period, to obtain the next spectral intensity. The training and testing were performed on 390 and 168 propagations respectively. The splitting of the testing and training data was performed to ensure that the testing data was a representative sample of the overall dataset. The predictions by the RNN $\hat{x}$ are compared with the calculated  values $x$ from the UPPE model, via the normalised root mean square (RMS) error defined as,
\begin{equation}
    R = \sqrt{\frac{\sum_{\omega,z} (x_{\omega,z} - \hat{x}_{\omega,z})^2}{\sum_{\omega,z} (x_{\omega,z})^2}},
\end{equation}
which is equivalent to the relative error, with $\omega$ and $z$  being the spectral and spatial coordinate respectively. The simulations can be fed with an arbitrary sampling for both the spectral and spatial axes, because the network usually does not require the extremely fine scanning used by the traditional  numerical simulations. The implemented sampling has then been determined via a heuristic approach since a high resolution would lead to higher accuracy on the expense of slowing down the performance of the trained network. For this example, we have opted for a 500$\times$150 points grid, which results in a 15 nm spectral resolution and a 30 $\mu$m spatial resolution. The Dense output layer has been assigned a \textit{sigmoid} activation function, since the training data has been normalised within the range (0,1).

\begin{figure}
    \centering
    \includegraphics[width=11.1cm]{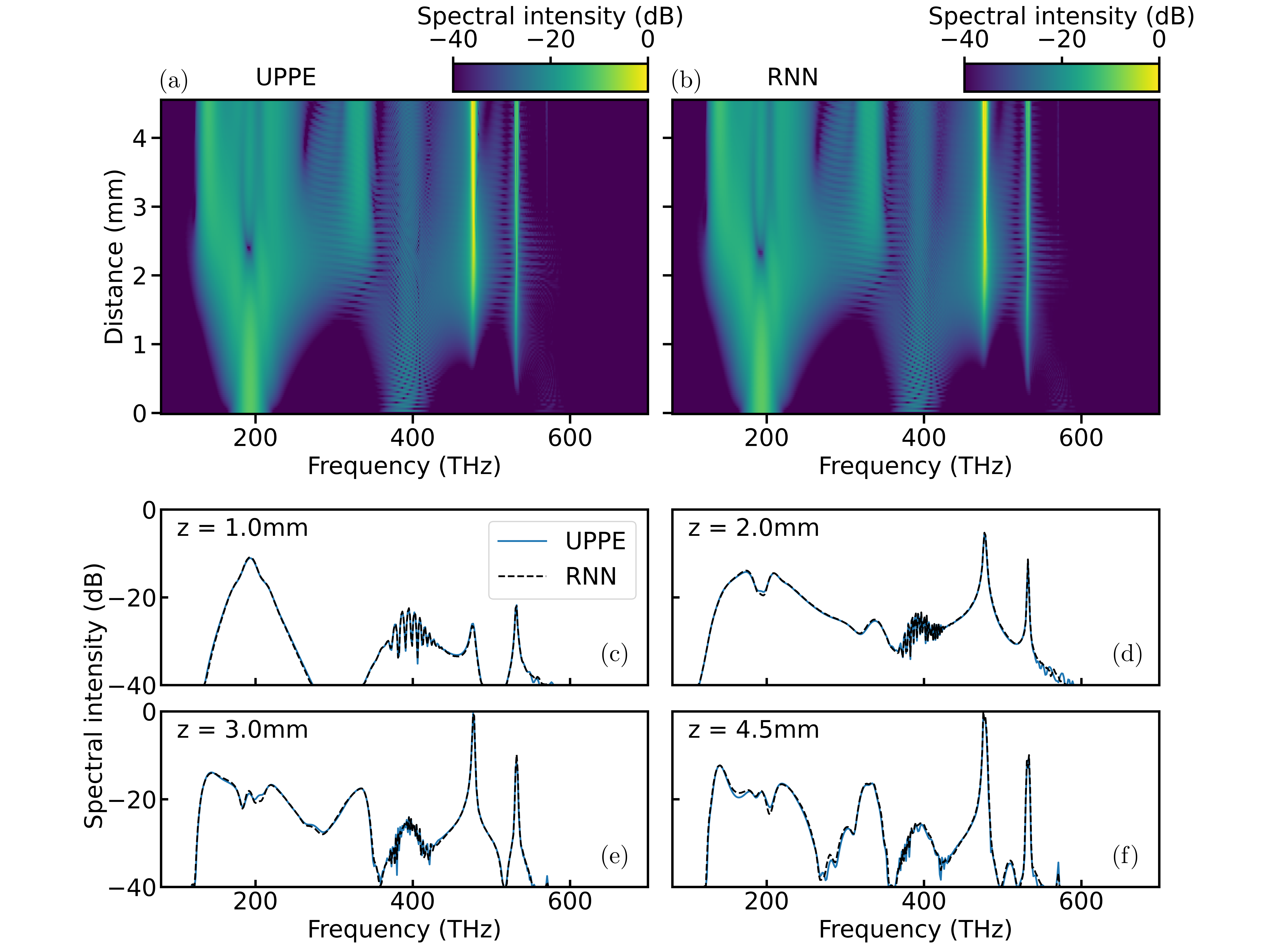}
    \caption{(a,b) Spectral evolution of an optical pulse centred at 1.55 $\mu$m with input energy 3.75 pJ and temporal FWHM 28.5 fs, in a 4.56 mm-long LiNbO$_3$ waveguide with a uniform poling period $\Lambda$ = 3.9 $\mu$m, predicted by the numerical UPPE model (a) and conditional neural network (b). (c-f) Spectral intensity profiles at four different positions along the waveguide.}
    \label{fig:Spectral1}
\end{figure}
\begin{figure}
    \centering
    \includegraphics[width=11.1cm]{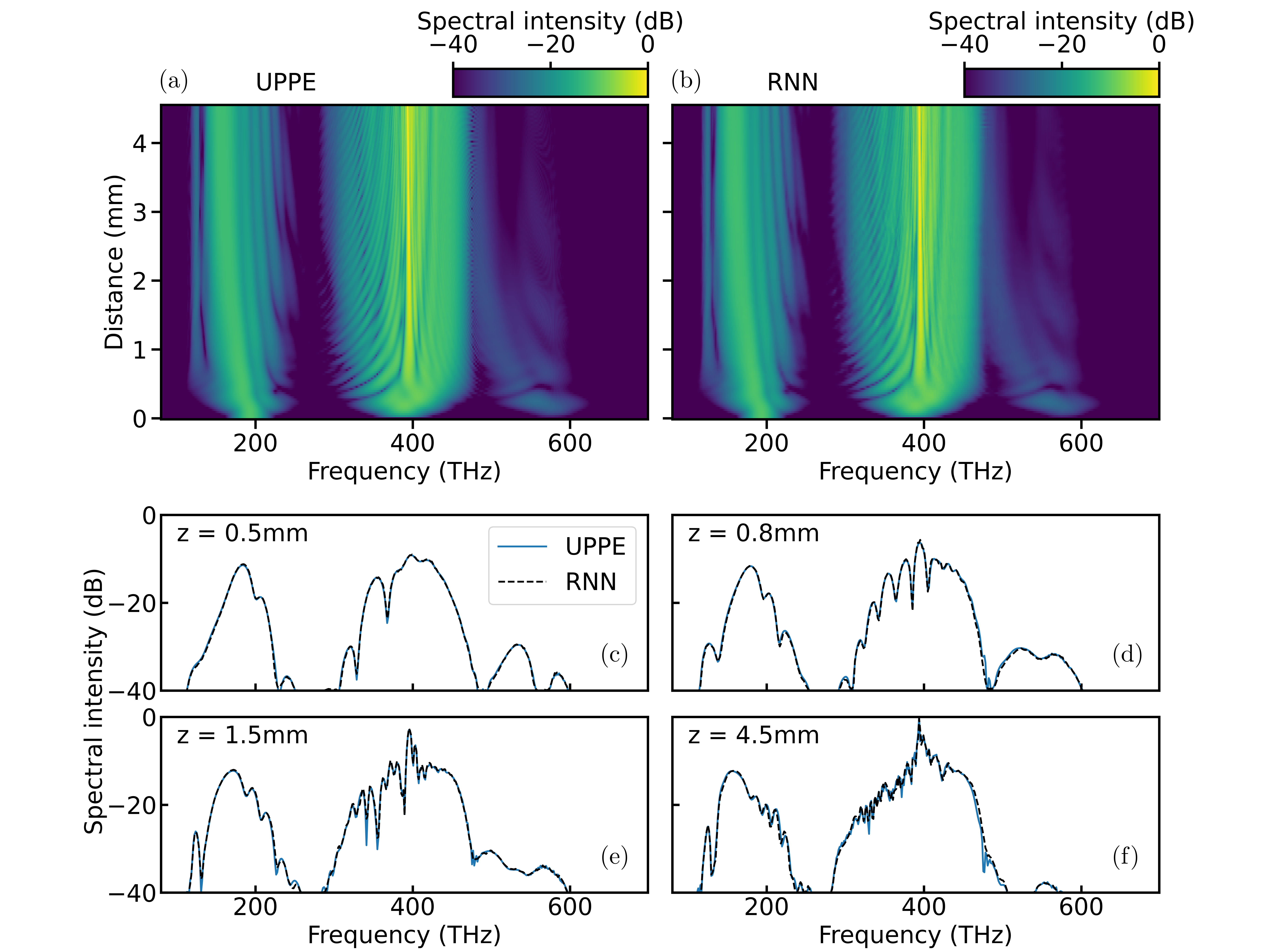}
    \caption{(a,b) Spectral evolution of an optical pulse centred at 1.55 $\mu$m with input energy 4.25 pJ and temporal FWHM 26.5 fs, in a 4.56 mm-long LiNbO$_3$ waveguide with a uniform poling period $\Lambda$ = 4.2 $\mu$m, predicted by the numerical UPPE model (a) and conditional neural network (b). (c-f) Spectral intensity profiles at four different positions along the waveguide.}
    \label{fig:Spectral2}
\end{figure}

Figures \ref{fig:Spectral1} and \ref{fig:Spectral2} compare the spectral evolution predicted by the UPPE model and the conditional RNN for two different poling periods $\Lambda$ = 3.9 $\mu$m and 4.2 $\mu$m, respectively, as well as different sets of pulse parameters that have not been used in the training process. A comparison between the results of the two models at four specific locations along the waveguide is also depicted. As portrayed, changing the poling period has dramatically modified the pulse dynamics inside the waveguide \cite{lauria}. Nevertheless, the outcomes of the UPPE and the conditional RNN are in excellent agreement in both cases. This is reflected in the normalised RMS errors being R = 0.015 and 0.019, for the former and latter cases, respectively. The predictions are also very good across  all the testing data, with very small fluctuation in accuracy.

The same architecture displayed in Fig. \ref{fig:Single} is also trained for the prediction of pulses through linearly-chirped poled waveguides that can provide quasi phase-matching for multiple second-order nonlinear interactions. The period at any point along the propagation axis $z$ is given by,
\begin{equation}
\frac{2 \pi}{ \Lambda (z)} = \frac{2\pi}{\Lambda _0} + \kappa z,
\end{equation}
where $\kappa$ is the chirp parameter that depends on the initial $\Lambda_0$ and final $\Lambda_L$ poling periods. The shape of the continuous chirp function can be discretised as a stair-like function with multiple  segments, each of constant poling period \cite{lauria}. The training is performed over a 10-mm long lithium niobate waveguide, where the poling-period at each step is set as the conditional parameter. The  pulse energy and temporal  duration ranges are 8 -- 12 pJ and 30 -- 40 fs, corresponding to dispersion and nonlinearity lengths in the ranges 7.59--13.5 mm and 0.41--0.82 mm, respectively. The linearly-chirped poling period varies within the domain 3.8 -- 4.2 $\mu$m. A lower spatial resolution of 50 $\mu$m has been implemented, while maintaining the same spectral resolution, resulting in a 500$\times$200 points grid. The model was trained on 60 propagations and tested on 30, with both subsets being a representative sample of the entire dataset. The rest of the simulation and training parameters are the same as in the uniform poling case.

\begin{figure}[H]
    \centering
    \includegraphics[width=11.1 cm]{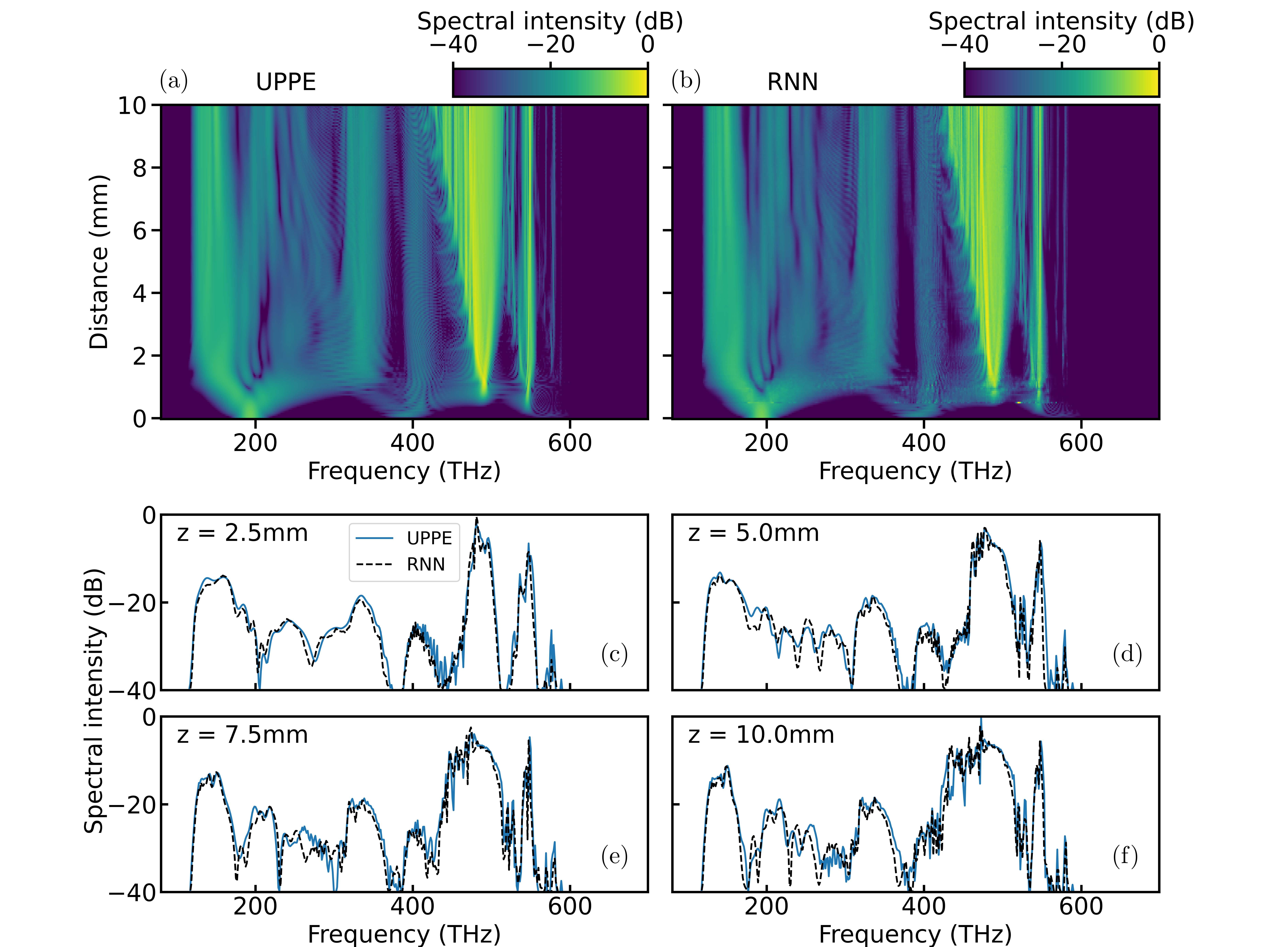}
    \caption{(a,b) Spectral evolution of a pulse of an input energy 12 pJ and temporal width 30 fs in a 10 mm-long LiNbO$_3$ waveguide with a chirp poling period from $\Lambda_0$ = 3.85 $\mu$m to $\Lambda_L$ = 4.1 $\mu$m, predicted by the UPPE model (left) and conditional neural network (right). (c-f) Spectral intensity at different distances throughout the propagation.}
    \label{fig:Chirp}
\end{figure}

The spectral evolution of an optical pulse with parameters within the aforementioned ranges in a waveguide with a poling period in the range 3.85 -- 4.1 $\mu$m  using the UPPE and conditional RNN are portrayed in Fig. \ref{fig:Chirp}. The shown simulations are for a set of parameters belongs to the testing data. As depicted, a very good agreement between the two models has been obtained, with a normalised RMS error of R = 0.08. 

Similar results have also been obtained for other linearly-chirped waveguides with different poling-period ranges.
Although the sizes of the training and testing datasets are relatively small in comparison to the previous application, because of the associated training time needed in this case, a very good agreement is still obtained  across the whole testing dataset.

\section{Double conditional RNN}
\begin{figure}[H]
    \centering
    \includegraphics[width=0.9\textwidth]{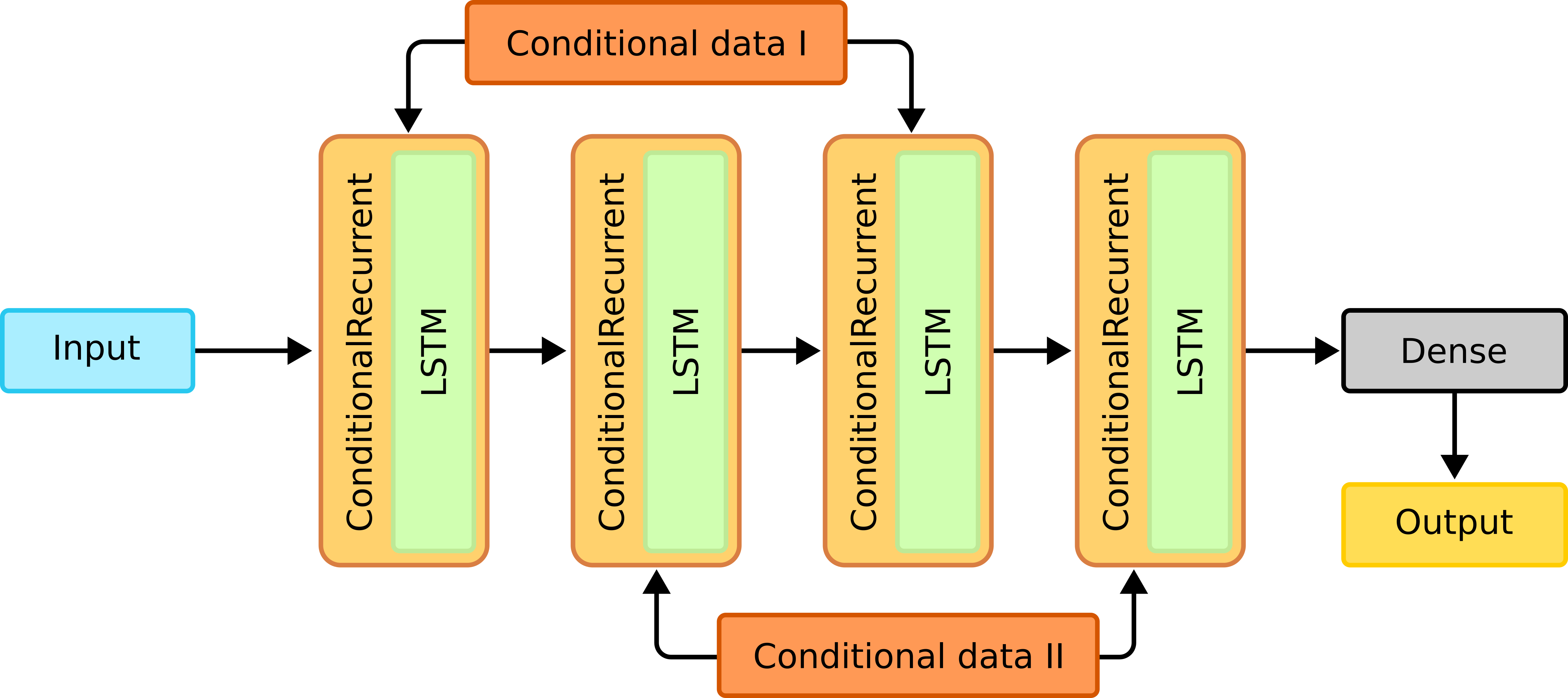}
    \caption{Architecture of a dual conditional model, where each conditional data is fed twice to the network and the sequential data is  fed only once.}
    \label{fig:Dual}
\end{figure}

The network can also be trained to predict the real and imaginary parts of the complex envelope $\mathcal{A} $, by setting them as a boolean switch, to inform the network if the prediction being performed corresponds to the real or imaginary part. Hence, both the magnitude and phase of the complex envelope $\mathcal{A} $ can be determined using the same network. Subsequently, predicting the pulse spectral evolution would allow us to calculate its temporal evolution via applying the inverse Fourier Transform, provided the data is on a Fourier grid. The ability of the network to retain the phase of the complex envelope widens the applications of the proposed technique, such as in calculating the spectral coherence and the cross-frequency resolved optical gating (XFROG). Moreover, this condition can be combined together with the poling period in a double-condition recurrent network, as shown in Fig. \ref{fig:Dual}. Similarly to the single-conditional network, each condition needs to be fed twice for the best initialisation of the hidden states. 
The conditional RNN  is optimised and designed for a single condition. Therefore, in the case of having multiple conditions, each layer should be initialised for each condition individually.  We found  that either feeding the two conditions simultaneously or feeding the first condition twice followed by the second condition twice and so on would result in a less accurate prediction.

Using the simulations generated by the UPPE model, the architecture displayed in Fig. \ref{fig:Dual} has been trained on the spectral evolution of the real and imaginary parts of the pulse complex envelope in uniformly-poled waveguides, where the period is in the range 3.4 -- 3.8 $\mu$m. The simulation parameters are the same as the previous examples except for the pulse energy and temporal duration, that are varied over the ranges 5 -- 7 pJ and 25 -- 40 fs, respectively. The model was trained and tested on 88 and 44 propagations respectively, and the subsets were selected to both be a representative sample of the entire dataset. The sequence length is set to $N = 10$, and the two conditions, the poling period and the complex envelope, are fed at each step. The data is sampled with a spectral resolution of $\simeq$ 1.5 $\mu$m to allow the application of the inverse Fourier Transform to correctly restore the pulse temporal evolution. The number of points along the frequency axis is still reduced to $\sim$ 5k by removing the points outside the transparency window of the waveguide, which are added back as zeros after the prediction is performed by the network. We found that this would have a negligible effect on the complex envelope. The spatial resolution is set at 30 $\mu$m. The Dense output layer has been set to a \textit{tanh} activation function, since the data  in this case is normalised in the range (-1,1). 

\begin{figure}
         \centering
    \includegraphics[width=11.1cm]{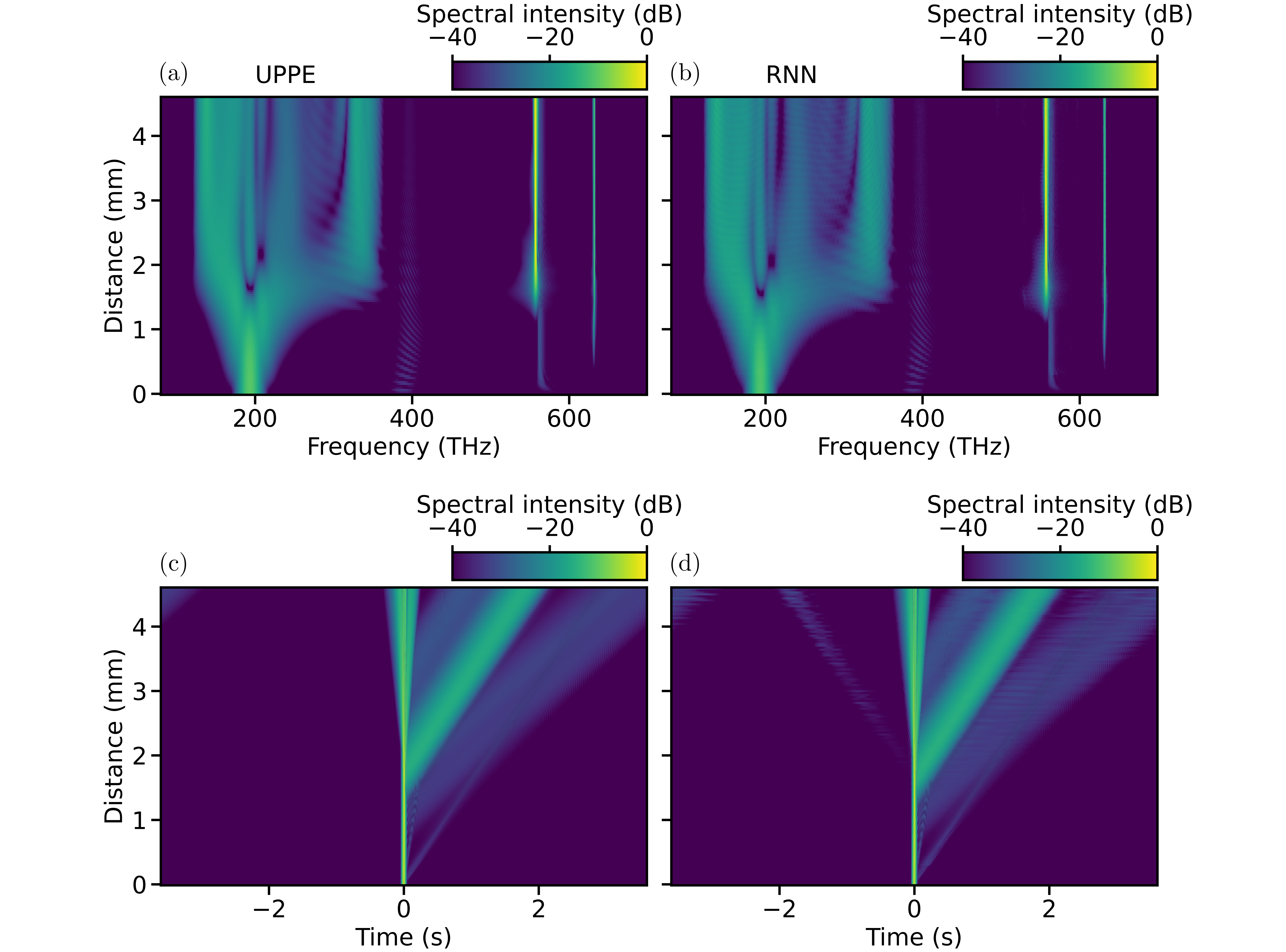}
	\caption{(a,b) Spectral and (c,d) temporal evolution of a pulse of an input energy 6 pJ  and temporal width  30 fs in a 4.56 mm-long LiNbO$_3$ waveguide with uniform poling period $\Lambda$ = 3.4 $\mu$m, using the UPPE model (a,c) and the double conditional neural network(b,d).}
	\label{fig:ComplexDual}
\end{figure}

The spectral and temporal evolution of an optical pulse with an input energy 6 pJ and temporal width 30 fs in a 4.56 mm-long lithium niobate waveguide with a uniform poling period $\Lambda$ = 3.4 $\mu$m is displayed in Fig. \ref{fig:ComplexDual}. The pulse parameters are not involved in the training process. As shown, the results by the RNN and UPPE models are in very good agreement, with a normalised RMS error of R = 0.009 and R = 0.03 for the spectral and temporal evolution, respectively.

\section{Conclusions}
To summarise and compare the performance of the presented conditional RNNs and the UPPE model, Table 1 shows the simulation parameters and speed of the different approaches described in this work. Since the waveguide lengths can vary, we have compared the average simulations speed rather than the time. The simulation speed for the UPPE is an estimate, since the adaptive algorithm can significantly slow down the simulations to avoid divergence.  As illustrated, the conditional RNN can be up to 60 times faster than the UPPE model for the same dynamics, demonstrating its potential in predicting the outputs of nonlinear pulse propagation in waveguides. The average normalised RMS errors for each application are also shown in the table. They have been calculated over the entire testing sets for each application, and they show good agreement between the RNN predictions and the numerical solution results. Also, the obtained values of RMS error  are  better than other related works \cite{gentyRNN, gentyFFNN}, that achieve values  within the range 0.09--0.19.

\begin{table}[H]
\caption{Comparison between the UPPE and  trained NNs used in Figs. \ref{fig:Spectral1},\ref{fig:Spectral2} (Uniform NN),  Fig. \ref{fig:Chirp} (Chirp NN), and Fig. \ref{fig:ComplexDual} (Complex NN) in terms of the simulation parameters and speed.}
\begin{tabular}{lllll}
\hline
                          & UPPE              & Uniform NN    & Chirp NN      & Complex NN     \\ \hline
RMS                       & N/A               & 0.016         & 0.088         & 0.011 (0.04$^a$)  \\ \hline
Number of spectral points & \textgreater{}32k & 500           & 500           & $\sim$5k       \\ \hline
Spectral resolution       & 0.14 THz          & 1.5 THz       & 1.5 THz       & 0.14 THz       \\ \hline
Longitudinal stepsize     & 0.2 $\mu$m$^b$     & 30 $\mu$m     & 50  $\mu$m    & 30  $\mu$m     \\ \hline
Simulation speed          & 0.03 mm/s         & 0.9 mm/s      & 1.6 mm/s      & 0.2 mm/s       \\ \hline
Training time$^c$          & N/A               & $\sim$1.7 hrs & $\sim$3.5 hrs & $\sim$15.6 hrs \\ \hline
Network variables         & N/A               & 4M            & 4M            & 57M            \\ \hline
\end{tabular}
\caption*{$^a$ Temporal evolution predicted indirectly from spectral envelope using the inverse Fourier Transform. $^b$ Step-size usually becomes much shorter than this value via the implemented adaptive algorithm, particularly in the pulse compression regime. \\$^c$ Average time to train with 200 propagations for 500 epochs.}

\end{table}

In conclusion, we have exploited the single and dual conditional architectures within LSTM models in predicting nonlinear pulse propagation in optical periodically-poled nanowaveguides with second and third-order nonlinearities. These innovative approaches have demonstrated their capability to significantly outpace traditional numerical models while maintaining remarkable accuracy.  The successful integration of conditional data into sequential data is a critical breakthrough in pulse-propagation modelling. Notably, this work marks the first success in predicting both the spectral and temporal evolution of a pulse using the same neural network, allowing for calculating the XFROG spectrograms or spectral coherence of the optical source.

This research will offer new opportunities for the development of faster and more efficient approaches to modeling nonlinear phenomena. In this work, we set the poling period and the real/imaginary parts of the complex envelope as conditions for the RNNs. However, other conditional parameters can be included and explored as well, such as the waveguide geometry or different core materials. Since the proposed approach is fully data-driven, it can be easily applied for other nonlinear physical systems, for instance, Bose-Einstein condensates that can be described by the Gross–Pitaevskii equation \cite{bose_ein}. Since the proposed networks rely on `Supervising Learning', they are unable to predict phenomena not captured by the model used in training. However, the RNNs perform the predictions at a constant rate, regardless  the complexity of the dynamics that can dramatically reduce the computational speed. For instance, when the cascaded second-order nonlinearity \cite{cascaded} is dominant, we found that the computational time of traditional models can reach an hour or more for just few millimeters-long propagation, whereas the RNN models always predict the outcome in few seconds. This approach can hence be utilised in optimising a  wide range of input parameters to tailor a certain output. Finally, we envisage that this work would offer fruitful opportunities for expanding the applications of machine learning in linear and nonlinear optics.

\begin{backmatter}
\bmsection{Acknowledgements} This research is supported by EPSRC Doctoral Training Partnerships (DTP) programme.

\bmsection{Disclosures}
The authors declare no conflict of interest.

\bmsection{Data availability}
Data underlying the results presented in this paper are not publicly available at this time but may be obtained from the authors upon reasonable request.
\end{backmatter}
\bibliography{sample}

\begin{thebibliography}{10}
\newcommand{\enquote}[1]{``#1''}

\bibitem{GeneralML}
I.~Goodfellow, Y.~Bengio, and A.~Courville, \emph{Deep Learning} (The Mit Press, 2016).

\bibitem{ML_trends}
M.~I. Jordan and T.~M. Mitchell, \enquote{Machine learning: Trends, perspectives, and prospects,} {\protect\JournalTitle{Science}} \textbf{349}, 255--260 (2015).

\bibitem{classification1}
A.~Soofi and A.~Awan, \enquote{Classification techniques in machine learning: Applications and issues,} {\protect\JournalTitle{Journal of Basic \& Applied Sciences}} \textbf{13}, 459--465 (2017).

\bibitem{calssification2}
S.~B. Kotsiantis, I.~D. Zaharakis, and P.~E. Pintelas, \enquote{Machine learning: a review of classification and combining techniques,} {\protect\JournalTitle{Artificial Intelligence Review}} \textbf{26}, 159--190 (2006).

\bibitem{NLP}
J.~Xiao and Z.~Zhou, \enquote{Research progress of rnn language model,}  (2020).

\bibitem{timeseries}
R.~P. Masini, M.~C. Medeiros, and E.~F. Mendes, \enquote{Machine learning advances for time series forecasting,} {\protect\JournalTitle{Journal of Economic Surveys}}  (2021).

\bibitem{cybersec}
A.~Handa, A.~Sharma, and S.~K. Shukla, \enquote{Machine learning in cybersecurity: A review,} {\protect\JournalTitle{WIREs Data Mining and Knowledge Discovery}} \textbf{9}, e1306 (2019).

\bibitem{healthcare}
R.~Bhardwaj, A.~R. Nambiar, and D.~Dutta, \enquote{A study of machine learning in healthcare,} in \emph{2017 IEEE 41st Annual Computer Software and Applications Conference (COMPSAC),}  vol.~2 (2017), pp. 236--241.

\bibitem{vehicle}
S.~Kuutti, R.~Bowden, Y.~Jin, P.~Barber, and S.~Fallah, \enquote{A survey of deep learning applications to autonomous vehicle control,} {\protect\JournalTitle{IEEE Transactions on Intelligent Transportation Systems}} \textbf{22}, 712--733 (2021).

\bibitem{neuroscience}
J.~I. Glaser, A.~S. Benjamin, R.~Farhoodi, and K.~P. Kording, \enquote{The roles of supervised machine learning in systems neuroscience,} {\protect\JournalTitle{Progress in Neurobiology}} \textbf{175}, 126--137 (2019).

\bibitem{NLML1}
G.~Genty, L.~Salmela, J.~M. Dudley, D.~Brunner, A.~Kokhanovskiy, S.~Kobtsev, and S.~K. Turitsyn, \enquote{Machine learning and applications in ultrafast photonics,} {\protect\JournalTitle{Nature Photonics}} \textbf{15}, 91--101 (2020).

\bibitem{NLML2}
D.~Piccinotti, K.~F. MacDonald, S.~Gregory, I.~Youngs, and N.~I. Zheludev, \enquote{Artificial intelligence for photonics and photonic materials,} {\protect\JournalTitle{Reports on Progress in Physics}} \textbf{84}, 012401--012401 (2020).

\bibitem{NLML3}
P.~Freire, E.~Manuylovich, J.~E. Prilepsky, and S.~K. Turitsyn, \enquote{Artificial neural networks for photonic applications---from algorithms to implementation: tutorial,} {\protect\JournalTitle{Adv. Opt. Photon.}} \textbf{15}, 739--834 (2023).

\bibitem{inverseDesign}
P.~R. Wiecha, A.~Arbouet, C.~Girard, and O.~L. Muskens, \enquote{Deep learning in nano-photonics: inverse design and beyond,} {\protect\JournalTitle{Photonics Research}} \textbf{9}, B182--B182 (2021).

\bibitem{microscopy}
V.~T. Hoang, Y.~Boussafa, L.~Sader, S.~Février, V.~Couderc, and B.~Wetzel, \enquote{Optimizing supercontinuum spectro-temporal properties by leveraging machine learning towards multi-photon microscopy,} {\protect\JournalTitle{Frontiers in photonics}} \textbf{3} (2022).

\bibitem{physicsNN}
X.~Jiang, D.~Wang, Q.~Fan, M.~Zhang, C.~Lu, and A.~P.~T. Lau, \enquote{Physics-informed neural network for nonlinear dynamics in fiber optics,} {\protect\JournalTitle{Laser \& Photonics Reviews}} \textbf{16}, 2100483 (2022).

\bibitem{gentyFFNN}
L.~Salmela, M.~Hary, M.~Mabed, A.~Foi, J.~M. Dudley, and G.~Genty, \enquote{Feed-forward neural network as nonlinear dynamics integrator for supercontinuum generation,} {\protect\JournalTitle{Optics Letters}} \textbf{47}, 802--802 (2022).

\bibitem{gentyRNN}
L.~Salmela, N.~Tsipinakis, A.~Foi, C.~Billet, J.~M. Dudley, and G.~Genty, \enquote{Predicting ultrafast nonlinear dynamics in fibre optics with a recurrent neural network,} {\protect\JournalTitle{Nature Machine Intelligence}} \textbf{3}, 344–354 (2021).

\bibitem{ConvNN}
H.~Sui, H.~Zhu, H.~Jia, Q.~Li, M.~Ou, B.~Luo, X.~Zou, and L.~Yan, \enquote{Predicting nonlinear multi-pulse propagation in optical fibers via a lightweight convolutional neural network,} {\protect\JournalTitle{Optics Letters}} \textbf{48}, 4889 (2023). Test.

\bibitem{NNaidNLSE}
R.-Q. Wang, L.~Ling, D.~Zeng, and B.-F. Feng, \enquote{A deep learning improved numerical method for the simulation of rogue waves of nonlinear schrödinger equation,} {\protect\JournalTitle{Communications in Nonlinear Science and Numerical Simulation}} \textbf{101}, 105896 (2021).

\bibitem{rnn_design}
L.~C. Jain, \emph{Recurrent neural networks : design and applications} (Crc Press, 2000).

\bibitem{UPPE}
M.~Kolesik and J.~V. Moloney, \enquote{Nonlinear optical pulse propagation simulation: From maxwell’s to unidirectional equations,} {\protect\JournalTitle{Physical Review E}} \textbf{70}, 036604 (2004).

\bibitem{lauria}
S.~Lauria and M.~F. Saleh, \enquote{Mixing second- and third-order nonlinear interactions in nanophotonic lithium-niobate waveguides,} {\protect\JournalTitle{Physical Review A}} \textbf{105} (2022).

\bibitem{Conforti13}
M.~Conforti, A.~Marini, T.~X. Tran, D.~Faccio, and F.~Biancalana, \enquote{Interaction between optical fields and their conjugates in nonlinear media,} {\protect\JournalTitle{Opt. Express}} \textbf{21}, 31239--31252 (2013).

\bibitem{Agrawal07}
G.~P. Agrawal, \emph{Nonlinear Fiber Optics}, San Diego, California (Academic Press, 2007), 4th ed.

\bibitem{rnns_time}
Z.~C. Lipton, J.~Berkowitz, and C.~Elkan, \enquote{A critical review of recurrent neural networks for sequence learning,}  (2015).

\bibitem{lstm}
S.~Hochreiter and J.~Schmidhuber, \emph{Long short term memory} (München Inst. Für Informatik, 1995).

\bibitem{vanish}
S.~Hochreiter, \enquote{The vanishing gradient problem during learning recurrent neural nets and problem solutions,} {\protect\JournalTitle{International Journal of Uncertainty, Fuzziness and Knowledge-Based Systems}} \textbf{06}, 107--116 (1998).

\bibitem{remy}
P.~Rémy, \enquote{Conditional recurrent for tensorflow/keras,}  (2023).

\bibitem{keras}
F.~Chollet \emph{et~al.}, \enquote{Keras,} \url{https://keras.io} (2015).

\bibitem{bose_ein}
M.~F. Saleh and P.~Öhberg, \enquote{Trapped bose–einstein condensates in the presence of a current nonlinearity,} {\protect\JournalTitle{Journal of Physics B: Atomic, Molecular and Optical Physics}} \textbf{51}, 045303 (2018).

\bibitem{cascaded}
M.~G. {Vazimali} and S.~{Fathpour}, \enquote{{Applications of thin-film lithium niobate in nonlinear integrated photonics},} {\protect\JournalTitle{Advanced Photonics}} \textbf{4}, 034001 (2022).

\end{thebibliography}

\end{document}